\documentclass[a4paper]{article}

\usepackage{INTERSPEECH2019}
\usepackage[final]{changes}

\title{Privacy-Preserving Speaker Recognition with Cohort Score Normalisation}

\name{Andreas Nautsch$^1$, Jose Patino$^1$, Amos Treiber$^2$, Themos Stafylakis$^3$, \\ Petr Mizera$^3$, Massimiliano Todisco$^1$, Thomas Schneider$^2$ and Nicholas Evans$^1$}
\address{
  $^1$Audio Security and Privacy, Digital Security Department, EURECOM, France\\
  $^2$Cryptography and Privacy Engineering, Department of Computer Science, TU Darmstadt, Germany\\
  $^3$Omilia -- Conversational Intelligence, Greece}
\email{\{nautsch,patino,todisco,evans\}@eurecom.fr, \{treiber,schneider\}@encrypto.cs.tu-darmstadt.de, \{tstafylakis,pmizera\}@omilia.com}

\usepackage{multirow}

\usepackage{todonotes}

\usepackage{subfig}

\usepackage{fontawesome}

\usepackage{numprint}

\usetikzlibrary{shapes,calc}
\makeatletter
\tikzset{
    database/.style={
        path picture={
            \draw (0, 1.5*\database@segmentheight) circle [x radius=\database@radius,y radius=\database@aspectratio*\database@radius];
            \draw (-\database@radius, 0.5*\database@segmentheight) arc [start angle=180,end angle=360,x radius=\database@radius, y radius=\database@aspectratio*\database@radius];
            \draw (-\database@radius,-0.5*\database@segmentheight) arc [start angle=180,end angle=360,x radius=\database@radius, y radius=\database@aspectratio*\database@radius];
            \draw (-\database@radius,1.5*\database@segmentheight) -- ++(0,-3*\database@segmentheight) arc [start angle=180,end angle=360,x radius=\database@radius, y radius=\database@aspectratio*\database@radius] -- ++(0,3*\database@segmentheight);
        },
        minimum width=2*\database@radius + \pgflinewidth,
        minimum height=3*\database@segmentheight + 2*\database@aspectratio*\database@radius + \pgflinewidth,
    },
    database segment height/.store in=\database@segmentheight,
    database radius/.store in=\database@radius,
    database aspect ratio/.store in=\database@aspectratio,
    database segment height=0.1cm,
    database radius=0.25cm,
    database aspect ratio=0.35,
}
\makeatother

\usepackage{placeins} 
    
\begin{document}

\maketitle

\begin{abstract}

In many voice biometrics applications there is a requirement to preserve privacy, not least because of the recently enforced General Data Protection Regulation (GDPR). Though progress in bringing privacy preservation to voice biometrics is lagging behind developments in other biometrics communities, recent years have seen rapid progress, with secure computation mechanisms such as homomorphic encryption being applied successfully to speaker recognition. Even so, the computational overhead incurred by processing speech data in the encrypted domain is substantial. While still tolerable for single biometric comparisons, most state-of-the-art systems perform some form of cohort-based score normalisation, requiring {\it many thousands of} biometric comparisons. The computational overhead is then prohibitive, meaning that one must accept either degraded performance (no score normalisation) or potential for privacy violations. This paper proposes the first computationally feasible approach to privacy-preserving cohort score normalisation. Our solution is a cohort pruning scheme based on secure multi-party computation which enables privacy-preserving score normalisation using probabilistic linear discriminant analysis (PLDA) comparisons. The solution operates upon binary voice representations. While the binarisation is lossy in biometric \mbox{rank-1} performance, it supports computationally-feasible biometric \mbox{rank-n} comparisons in the encrypted domain.

\end{abstract}

\section{Introduction}

Today there is a growing drive to bring privacy preservation to the realm of speech processing.  
Following new privacy regulation such as the European GDPR~\cite{EU-GDPR}, technology to protect sensitive data, including voice data, is attracting the attention of researchers and industrial stakeholders alike.  
Perhaps the most compelling argument to preserve privacy in speech signals is because they represent inherently personal and private information.  Examples include paralinguistic and extralinguistic information, attributes and characteristics, e.g., gender, age, language, dialect, accent, health status, general well-being and emotional state---and the biometric identity. 





This paper concerns the protection of privacy for voice biometric applications, e.g.~speaker recognition.  Recent years have seen rapid progress in privacy-preserving speaker recognition, e.g.~\cite{portelo2014gc,Paulini2016Odyssey}.  The most recent contribution to the field~\cite{nautsch2018homomorphic} reported the first
i-vector-based solution using homomorphic encryption (HE).  HE supports computation upon sensitive biometric voice data \emph{in the encrypted domain} and is a popular tool for
privacy preservation.  However, the computational demands of HE are prohibitive.  This is especially true in the case of speaker recognition systems that employ some form of cohort score normalisation.  When operating in unconstrained environments cohort score normalisation is key to performance and is a feature of any state-of-the-art solution.  Unfortunately, cohort score normalisation only compounds the computational burden of encryption since it typically involves many thousands of biometric comparisons in the scoring of a single utterance.  The scale of the computational demands are currently a bottleneck to privacy preservation for speaker recognition.

The work reported in this paper aims to overcome this bottleneck with an alternative, efficient approach to cohort score normalisation.  
Using an efficient approach to speaker modelling~\cite{Anguera2010Interspeech}, we propose to replace the speaker representation used in cohort score normalisation with an alternative \emph{binary key} (BK) representation.  As a native binary representation, BKs are readily suited to efficient computation in the encrypted domain. 
The paper shows that the computational overhead of operating upon encrypted representations can then be reduced greatly, meaning that probabilistic linear discriminant analysis (PLDA) comparisons can, for the first time, be performed in the encrypted domain with realistic computational resources. 

This paper is organised as follows.
Section~\ref{sec:secure-computation} describes the related work in privacy-preserving speaker recognition.
Section~\ref{sec:binary-keys} describes BK voice representations. Section~\ref{sec:cohort-pruning-cyptosystem} describes the proposed efficient cohort pruning scheme using BK representations and shows how it can be employed for privacy-preserving score normalisation.
Section~\ref{sec:experiments} presents an experimental validation. 
Conclusions are provided in Section~\ref{sec:conclusion}.

\section{Preliminaries and Related Work}
\label{sec:secure-computation}

There is an extensive body of literature concerning the preservation of privacy in biometrics.  Unfortunately, most relates not to speaker recognition, but to other biometric characteristics, e.g. fingerprint, iris, and face recognition~\cite{blanton2011secure, bringer2014gshade}.  Whatever the characteristic, the requirements for effective privacy preservation are the same.  These are outlined in the ISO/IEC~24745 standard~\cite{ISO-IEC-24745-TemplateProtection-101129} which stipulates that biometric information must be \emph{unlinkable} (data of protected databases are not relatable), \emph{irreversible} (neither embeddings nor audio can be recreated from protected data), and \emph{renewable} \replaced{(no biometric voice data needs to be recaptured to update a privacy-preservation algorithm)}{(when updating a privacy-preservation algorithm, no biometric voice data needs to be recaptured)}.

The conventional approach to meet these requirements involves some form of encryption.  Since the late 1980s, the focus of the cryptographic community is  \emph{secure computation}~\cite{yao1982protocols, goldreich1987play}, specifically the evaluation of a function in ways that do not reveal any information about the inputs of the involved parties, except for the results.  Secure computation mechanisms may be harnessed to retain the functionality of an application without compromising the privacy of the involved parties.
The main techniques include homomorphic encryption (HE) which enables computations to be carried out on ciphertexts,
and secure multi-party computation (SMPC) which allows interactive computations on data that is \emph{secretly shared}\footnote{E.g., in the Boolean \replaced{Goldreich-Micali-Wigderson (GMW)}{GMW} protocol~\cite{goldreich1987play} for two parties that we will use in our work, an input bitstring $x$ can be secretly shared among the parties by sending a random bitstring $r$ of the same length to one party and sending $x \oplus r$ to the other party. Then, the GMW protocol can be executed to securely compute any functionality on $x$ using just the shares of $x$. The inputs stay hidden because neither $r$ nor $x \oplus r$ reveal any information about $x$.} between the parties.\footnote{
    Depending on the use case, a party could be a client device, an authentication, or a database/processing server. In contrast to the plaintext domain (one party is sufficient to carry out a computation), security in SMPC is established by splitting computations in a distributed system architecture, where each party computes only on secretly shared data.
}

Recent advances in state-of-the-art implementations of secure computation protocols (cf.~\cite{hastings2019sok}) have shown to be efficient solutions to privacy preservation in a wide variety of applications~\cite{demmler2015}.  
Even so, different solutions offer different levels of computational complexity. SMPC protocols typically involve multiple rounds of interaction (communications between parties involved in the secure computation).
While not necessarily requiring interaction, HE usually incurs a higher computational overhead. It follows that, while deployed secure computation techniques can be highly efficient and scalable, it depends on the use case and the employed mechanisms.

Both SMPC and HE have been applied successfully to privacy-preserving speaker recognition~\cite{portelo2014gc, smaragdis2007framework, pathak2013, aliasgari_secure_2013, aliasgari2017secure}. This body of work explores privacy preservation in traditional Gaussian mixture model (GMM) and hidden Markov model (HMM) architectures.  Typically, HE is used to hide biometric information, while scoring is sometimes performed using SMPC. The solution reported in~\cite{aliasgari_secure_2013, aliasgari2017secure} preserves privacy in an HMM framework by storing the corresponding secret shares among multiple servers, a technique known as outsourced SMPC~\cite{kamara2011secure}.
Of course, software solutions are not the only approach to privacy preservation.
The work in~\cite{brasser2018voiceguard} shows how privacy can also be preserved by using trusted execution environments such as the Intel SGX architecture~\cite{mckeen2013sgx}.

Recently, in~\cite{nautsch2018homomorphic}, an HE-based solution to privacy preservation in the form of the Paillier cryptosystem has been applied to state-of-the-art speaker recognition architectures including i-vector systems using PLDA.  This work shows that a one-to-one PLDA comparison can be computed in a few hundred milliseconds, depending on whether the speaker model is also protected.  Unfortunately, while the solution delivers privacy preservation with no degradation to computational precision, it does not scale well.
Protection of a cohort score normalisation process which requires many thousands of comparisons is computationally prohibitive; a runtime in the order of $50$ minutes would be needed to process one reference-probe comparison involving \numprint{10000} cohort comparisons, a representative number for today's state-of-the-art techniques.

With cohort score normalisation being a feature of any state-of-the-art approach to speaker recognition, and with performance degradation being the cost of its omission, there is hence an interest to devise computationally manageable solutions.  With no previous work having considered this problem thus far, this is the goal of the research reported in this paper.
\section{Binary Key Voice Representations}\label{sec:binary-keys}

Binary voice representations have been reported previously in the context of privacy preservation.  Cryptobiometric (extraction/binding of cryptographic keys \emph{from} biometric data)\footnote{
    In contrast, HE uses cryptographic keys \emph{for de-/encrypting} biometric data (\emph{biometrics in the encrypted domain}).
} systems based upon the binarisation\footnote{
    The term \emph{binarisation} is potentially misleading. It refers to a \emph{higher level} binary representation (under the acceptance of precision loss) of digital speech data (which is itself already stored in \emph{binary} bit form).
} of GMM-based supervectors are reported in~\cite{Billeb2015IET,Paulini2016Odyssey}.  
%
The work in this paper uses an alternative, more elaborate approach based upon \emph{binary keys} (BKs), originally proposed in~\cite{Anguera2010Interspeech,bonastre2011discriminant}.  The BK approach takes a more speaker-discriminatory approach to modelling, much like the idea behind \emph{anchor models}~\cite{merlin1999non,mami2002speaker}. The same versatile approach has been applied successfully to a number of related problems including emotion recognition~\cite{luque2014modeling}, speaker change detection~\cite{patino2017speaker}, speaker diarization~\cite{delgado2015fast, Patino2018Interspeech} and privacy preservation~\cite{mtibaa2018cancelable} in the context of \emph{cancelable biometric systems} (irreversible feature transforms). Full details of the implementation used in this paper can be found in~\cite{Anguera2010Interspeech}.


\begin{figure}[!t]
    \centering
    \includegraphics[]{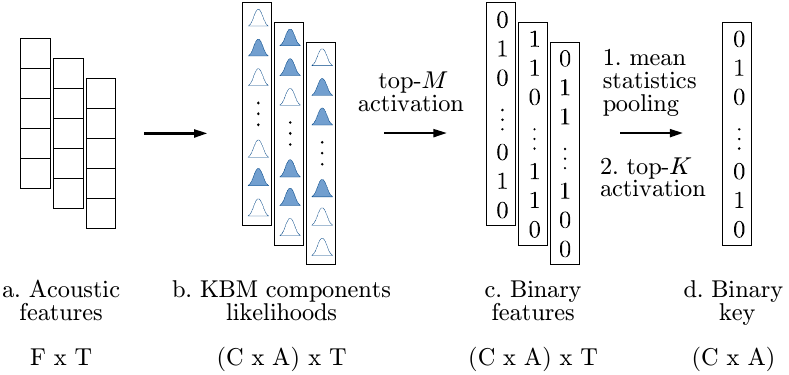}
    \caption{BK extraction process from $T$ frames with $F$-dimensional acoustic features to BKs from a KBM with $A$ anchors for each of the $C$ UBM components. Before setting $K$ KBM elements as \emph{True} at the sample level, $M$ elements are pre-selected at the frame level.}
    \label{fig:bk_extraction}
    \vspace{-0.5cm}
\end{figure}

The extraction of BKs is performed using a so-called binary key background model~(KBM).  The KBM plays a role similar to that of a conventional universal background model~(UBM) but, instead of representing the acoustic space in an expected sense, it is formed from the concatenation of a number $A$ of speaker-dependent models (learned using traditional UBM maximum a-posteriori adaptation). The role of the KBM anchors is similar in principle to a latent speaker subspace (PLDA alike; in a rough approximation), \replaced{namely}{hence} the extraction of discriminant BKs.

The BK extraction process is illustrated in Figure~\ref{fig:bk_extraction}. 
From acoustic features (a), KBM component likelihoods are computed (b). Similarly to i-vector extraction~\cite{dehak2011front}, in which component posteriors are pooled to zero order statistics, top-$M$ likelihoods (which at the frame level equals the top-$M$ component posteriors) are used to determine the most frequently activated components (c); again, a rough approximation. An even more compressed speaker representation (d) is obtained with the final BK representation which indicates simply the $K$ elements with the highest pooled mean statistics.

The research hypothesis under investigation is: the loss in precision will be tolerable given their use only for cohort pruning; their use will cause only marginal degradation to the benefit of score normalisation while nonetheless facilitating privacy.  


\section{Privacy-Preserving Cohort Pruning}
\label{sec:cohort-pruning-cyptosystem}

The contribution in this paper is an efficient, privacy-preserving approach to score normalisation. It is based upon  cohort pruning using BK speaker representations that allow for efficient computation in the encrypted domain. \added{The use of HE-protected i-vectors here is too slow; unprotected i-vectors are not \emph{unlinkable}.}  The following describes the approach and shows how computation is performed using SMPC while preserving the privacy of both data subjects and cohort speakers.

\subsection{Score normalisation}

Score normalisation is a processing step of any state-of-the-art approach to ASV.  It is applied to remove nuisance bias and variation that would otherwise influence comparison scores in diverse environmental conditions.  The general approach to normalisation is based upon a set of auxiliary scores resulting from comparisons between references, probes, and cohort data.  
A score $S$ is normalised to $S'$ according to $S' = \frac{S - \mu}{\sigma}$, where the mean $\mu$ and standard deviation $\sigma$ are derived from (Gaussian distributed) scores of comparisons with cohort data. 

In the case that comparisons involve reference data, this approach is referred to as zero normalisation (\emph{z-norm}).  In this case, cohort data characteristics are assumed to match those of the probe $\mathfrak{P}$ (which are fixed for one \emph{quality} condition). 
Normalisation is then performed using the mean $\mu_{\mathcal{R}}$ and standard deviation $\sigma_{\mathcal{R}}$ derived from the set of comparison scores $\mathcal{R}$.

Normalisation can also be applied using probe data.  This is known as test normalisation (\emph{t-norm}).  Here, cohort data characteristics are assumed to match those of the reference, in which case the cohort data consists of reference representations.  
Normalisation is then performed using the mean~$\mu_{\mathcal{P}}$ and standard deviation~$\sigma_{\mathcal{P}}$ derived from the set of comparison scores~$\mathcal{P}$.

Typically, cohort score distributions are rarely Gaussian-distributed.  
\emph{Adaptive} z-norm (az-norm) and t-norm (at-norm) are
commonly applied instead in order to account for this discrepancy. 
In practice, normalisation is performed with only the top-$n$ scores of $\mathcal{R}$ and $\mathcal{P}$ 
and, for i-vectors, both (a)z-norm and (a)t-norm are usually combined in symmetric fashion, giving (a)s-norm: $S' = \frac{1}{2}\left(\frac{S - \mu_{\mathcal{R}}}{\sigma_{\mathcal{R}}} + \frac{S - \mu_{\mathcal{P}}}{\sigma_{\mathcal{P}}}\right)$. The normalisation process too needs to preserve privacy. 

\subsection{Privacy preservation}
By using \cite{nautsch2018homomorphic}, privacy-preserving score normalisation can be performed using reference, probe, and cohort embeddings, all processed in the encrypted domain via HE-based PLDA (HE-PLDA)~\cite{nautsch2018homomorphic}.  The resulting, encrypted scores $\mathcal{R}$, $S$\added{,} and $\mathcal{P}$, none of which reveal any sensitive information, 
can then be decrypted by an authentication server in order that normalised scores $S'$ can be computed in the plaintext domain. 

Assessments of computing demands were performed with a Python implementation of HE-PLDA with two 400-dimensional embeddings, 64-bit floating point precision and a key size of 3072 bits (recommended by NIST~\cite{barker2016nist} in order to support adequate security given advances in computing power until 2030 and beyond) running on an Intel Core i9-7960X CPU  with 128 GB of RAM.  Computations require 320\,ms per comparison when only subject data is encrypted (target in this paper), and \numprint{973109}\,ms per comparison when both subject data \emph{and} PLDA model parameters are encrypted (the second architecture in~\cite{nautsch2018homomorphic}).  Since a cohort size exceeding some few thousand voice samples is not unusual, the privacy-preserving computation of $\mathcal{R},\mathcal{P}$ is computationally prohibitive.


\begin{figure}[!t]
    \centering
    \begin{tikzpicture}[font=\scriptsize]
        \node[rectangle,draw=black,align=center] (ref) {reference \\ sample};
        \node[rectangle,draw=black,align=center,right=8em of ref] (prb) {probe \\ sample};
        \node[rectangle,draw=black,align=center,below=1em of ref,xshift=3em] (refemb) {reference \\ embedding};
        \node[rectangle,draw=black,align=center,xshift=-3em] (prbemb) at (prb|-refemb) {probe \\ embedding};
        \node[rectangle,draw=black,align=center,left=1em of refemb] (refbk) {reference \\ binary key};
        \node[rectangle,draw=black,align=center,right=1em of prbemb] (prbbk) {probe \\ binary key};
        \node[database,align=center,below=1.5em of refemb,xshift=-7.5em,label=below:{cohort}] (dbcr) {};
        \node[below=1em of dbcr] {$\mathfrak{P}$ alike};
        \node[database,align=center,xshift=4.5em,label=below:{cohort}] (dbcp) at (prb|-dbcr) {};
        \node[below=1em of dbcp] {$\mathfrak{R}$ alike};
        \node[rounded corners,draw=black] (refbkand) at ([xshift=2.5em]refbk|-dbcr) {AND};
        \node[rounded corners,draw=black] (prbbkand) at ([xshift=-2.5em]prbbk|-dbcp) {AND};
        \node[draw=black,below=1em of refbkand,align=center] (tr) {top-n \\ pruning};
        \node[draw=black,align=center] (tp) at (prbbkand|-tr) {top-n \\ pruning};
        \node[rounded corners,draw=black] (PLDAcr) at ([xshift=2.5em,yshift=-7em]ref|-dbcr) {PLDA};
        \node[rounded corners,draw=black] (PLDAcp) at ([xshift=-2.5em]prb|-PLDAcr) {PLDA};
        \node[rounded corners,draw=black] (PLDA) at ($(PLDAcr)!0.5!(PLDAcp)$) {PLDA};
        \node[circle,draw=black,below=1em of PLDAcr,minimum width=2em] (Sr) {$\mathcal{R}$};
        \node[circle,draw=black,below=1em of PLDAcp,minimum width=2em] (Sp) {$\mathcal{P}$};
        \node[circle,draw=black,below=1em of PLDA,minimum width=2em] (S) {$S$};
        \node[rounded corners,draw=black,align=center,below=1em of S] (as) {as-norm};
        \draw[->] (ref) -- (refemb);
        \draw[->] (prb) -- (prbemb);
        \draw[->] (ref) -- (refbk);
        \draw[->] (prb) -- (prbbk);
        \draw[->] (dbcr) -- (refbkand);
        \draw[->] (refemb) -- ([xshift=1.5ex]PLDAcr.north);
        \draw[->] (refemb) -- (PLDA);
        \draw[->] (refbk) -- (refbkand);
        \draw[->] (prbbk) -- (prbbkand);
        \draw[->] (refbkand) -- (tr);
        \draw[->] (prbbkand) -- (tp);
        \draw[->] (dbcp) -- (prbbkand);
        \draw[->] (prbemb) -- ([xshift=-1.5ex]PLDAcp.north);
        \draw[->] (prbemb) -- (PLDA);
        \draw[->] (PLDAcr) -- (Sr);
        \draw[->] (PLDA) -- (S);
        \draw[->] (PLDAcp) -- (Sp);
        \draw[->] (Sr) -- (as);
        \draw[->] (S) -- (as);
        \draw[->] (Sp) -- (as);
        \draw[->] (tr) -- node (embnode) [pos=0.2,left] {embeddings} (PLDAcr);
        \draw[->] (tp) -- node (embnode2) [pos=0.2,right] {embeddings} (PLDAcp);
        %
        \draw[draw=red!80!black,densely dotted,ultra thick] ([xshift=-0.75em,yshift=-0.5em]PLDAcr.south west) rectangle ([xshift=0.75em,yshift=0.5em]PLDAcp.north east);
        \node[color=red!80!black,align=center] at ([xshift=3em]PLDAcp.east) {HE-PLDA,\\ using \cite{nautsch2018homomorphic}};
        \draw[draw=red!80!black,densely dotted,thick] ([xshift=-1.25em,yshift=-0.5em]dbcr.south west|-embnode) rectangle ([xshift=.85em,yshift=0.5em]refbkand.north east);
        \draw[draw=red!80!black,densely dotted,thick] ([xshift=-.5em,yshift=-0.5em]tp.south west|-embnode) rectangle ([xshift=1.25em,yshift=0.5em]dbcp.north east);
        \draw[draw=green!60!black, dashed,ultra thick] ([xshift=-1.5em,yshift=-0.75em]dbcr.south west|-embnode) rectangle ([xshift=1.5em,yshift=0.7em]dbcp.north east);
        \node[color=green!60!black] at ([yshift=0.75em]$(dbcr)!0.5!(dbcp)$) {proposed};
    \end{tikzpicture}
    \caption{Our proposed privacy-preserving as-norm protocol with cohort pruning (green dashed area). The red dotted areas indicate that operations are carried out in the encrypted domain and do not leak any information except the decryptable outputs.}
    \label{fig:big-picture-as-norm}
    \vspace{-0.5cm}
\end{figure}
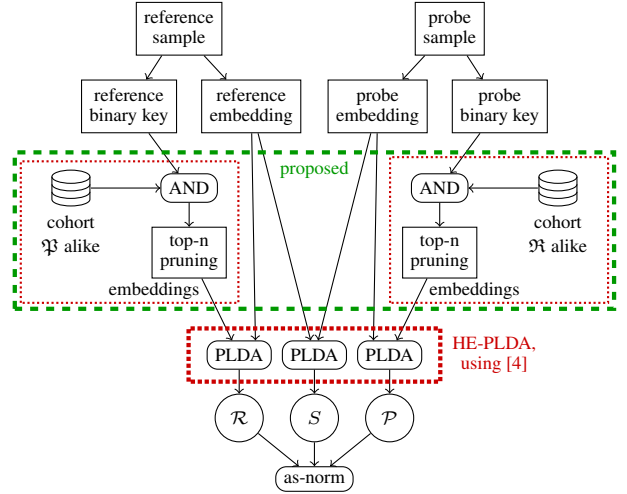

\subsection{Cohort pruning}

The research hypothesis under investigation here is that the selection of top-$n$ relevant cohort comparisons can be performed more efficiently by accepting some modest degradation to computational precision, while still preserving privacy.
Instead of selecting the top-$n$ cohort comparisons using ASV scores, they are selected \deleted{instead }using measures of acoustic similarity derived from BK representations of 
reference, probe, and cohort samples.  As illustrated in Figure~\ref{fig:big-picture-as-norm}, using BK representations, HE-PLDA can then be made efficient via secure bit-wise AND operations and top-$n$ cohort pruning.

More precisely, we employ the Boolean GMW protocol~\cite{goldreich1987play} (cf.\ Section~\ref{sec:secure-computation}) in the case of two involved parties to securely compute our proposed \deleted{secure }cohort pruning technique. Probe, reference\added{,} and cohort BKs are secretly shared between two servers that jointly and securely compute the top-$n$ pruning. 
This principle is \replaced{referred to as}{called} outsourced secure computation\deleted{ and was described in}~\cite{kamara2011secure}. 
Assuming non-colluding servers \replaced{as in}{and similarly to}~\cite{nautsch2018homomorphic}, this approach can tolerate the corruption of one server without any privacy leakage.
\replaced{This assumption}{The assumption of non-colluding servers} can be seen as realistic, given that one server could be supplied by an independent provider.
\added{Since we use protocols with \emph{semi-honest} security here, the secure pruning requires servers that honestly follow the protocol.}

Using the GMW protocol, we can easily compute an AND between the secretly shared sample and all secretly shared cohort data. On the resulting shares, we then securely compute the Hamming weight using the circuit of~\cite{boyar2005hamming} and perform a secure top-$n$ pruning as optimised in~\cite{JLLRSTY19}. As a result, the identifiers of the top-$n$ embeddings are revealed and can be used for score normalisation. Apart from this information, nothing else is leaked about the sample and cohort voice data.

\section{Experimental Validation}
\label{sec:experiments}



Given the research objective to demonstrate improvements in computational efficiency, rather than improved performance, only brief details of the text-independent speaker recognition system are provided here. It is based on 400-dimensional i-vectors, extracted from conventional acoustic features using time delay deep neural network (TDNN) for estimating UBM posteriors. The TDNN is trained using the KALDI toolkit~\cite{Povey_ASRU2011} with \mbox{SRE'04-08} and SWBD data\added{ (not x-vectors)}.  The \added{Python} backend is based on PLDA with mean and length normalisation, and trained with SRE'04-08 data.  The KBM used for BK extraction is learned \added{in Matlab} using a \numprint{2048}-component UBM trained with conventional acoustic features and $A=20$ anchors (10 fe/male each). 
KBM optimisation is performed using a subset of the cohort set containing data from 71 speakers. 
For BK extraction, at feature level the top $M=1$ components are activated, while at sample level the top $K=2048$ bits are set. 
The cohort set is a subset of the PLDA training set with \numprint{11640} voice samples of \numprint{3812} speakers. 
%
%
The proposed approach is evaluated \replaced{on}{using} the 2010 NIST SRE common condition\deleted{ (CC)} 5\deleted{ task}, particularly the core-core and core-10s \replaced{protocols}{conditions}. In order to report on diverse data, we \replaced{pooled}{mixed} the scores of both \replaced{protocols}{sets} (core-core/10s).


\subsection{Recognition results}

Results are reported in terms of $C_{\text{llr}}^{\text{min}}$, the minimum decision cost function (minDCF; effective prior 0.01)
and the equal-error rate (EER).
Table~\ref{tab:runtimes} shows that conventional 
as-norm gives the same or better performance than the baseline system (without any score normalisation). 
The proposed \emph{privacy-preserving} as-norm solution gives slightly worse results in terms of minDCF even though, curiously, improvements are observed in terms of $C_{\text{llr}}^{\text{min}}$ and EER.
For minDCF, an improvement over the baseline is also observed but without reaching the performance of the unprotected AS-norm. 
This result confirms our research hypothesis: privacy preservation incurs only a modest performance degradation (in the minDCF sense) and, encouragingly,
also improves upon the baseline system without any score normalisation (in the $C_{\text{llr}}^{\text{min}}$ and EER sense).

\subsection{Proof of biometric information protection}
The sample embeddings as well as the cohort embeddings used for PLDA comparisons are protected via the original privacy-preserving PLDA system. As such, if biometric information protection in the form of unlinkability, irreversibility, and renewability is given by the original system (as in~\cite{nautsch2018homomorphic}), then the embeddings are protected as well. The BKs of samples and cohorts are protected by the Boolean secret sharing of the GMW protocol between two servers (cf.\ Section~\ref{sec:secure-computation}). Because of the information-theoretic indistinguishability of any two secret shares, unlinkablitity and irreversibility are guaranteed. Due to the nature of secret sharing, the protected data is also renewable; secret shares can be re-randomised with a new random bitstring.

\subsection{Complexity analysis}
We implemented our secure cohort pruning architecture using the state-of-the-art SMPC framework \emph{ABY}~\cite{demmler2015}. We ran our implementations on two machines with Intel Core i9-7960X CPUs and 128 GBs of RAM. To simulate real-world network conditions of the involved servers, we restricted the connection between the servers to 1 Gbit/s bandwith and 1 ms round trip time. Results are presented in Table~\ref{tab:runtimes}. Note that these are the online runtimes and that some additional input-independent pre-computation is required; we account for the BK extraction time\footnote{
    BKs are extracted with Matlab on a DELL R620  with two Intel Xeon  E5-2630L CPUs and 128 GBs of RAM.
} (28.3\,s and 3.2\,s for a core-core and for a core-10s probe, respectively; for core-core/10s, the average is 16.9\,s). 
The largest gain in real-world network conditions for privacy-preserving score normalisation are observed for small cohorts with $14\times$ to $19\times$ gains in runtimes. In other words, rather than runtimes in the order of 50~minutes only a few minutes are necessary. In the privacy-preserving cohort pruning and as-norm, the BK extraction takes 6.4-20.0\% of the runtime and the GMW pruning takes 39.2-78.0\% (their time share is lower on higher cohort sizes as the runtime share of HE-PLDA increases). For the GMW pruning, all privacy-preserving az/at-norm comparisons (using BKs against the entire cohort) are carried out in less than 157\,s and 52\,s, respectively. These times already include the sorting of the top-50 cohort indices for pruning the HE-PLDA cohort comparisons. To prune larger cohort sizes, the privacy-preserving sorting requires additional time, e.g. from top-50 to top-400 in az-norm, an additional 126\,s are necessary.


\begin{table}[!t]
\caption{Runtimes and recognition results  
for the baseline system, the baseline system with conventional as-norm and the proposed privacy-preserving alternative, for different cohort sizes $n$. The realtime improvement results from dividing the time of scoring all cohort data with HE-PLDA by: BK extraction + GMW (scoring and top-$n$ sorting) + top-$n$ HE-PLDA time.}
\nprounddigits{0}\npdecimalsign{\ensuremath{.}}
\label{tab:runtimes}
\centering
\resizebox{80mm}{!}{
\begin{tabular}{r| r r r r r r r }
\toprule
$n$ & 50 & 100 & 150 & 200 & 250 & 300 & 400 \\
\midrule
Baseline ($C_{\text{llr}}^{\text{min}}$ / minDCF / EER) & \multicolumn{7}{c}{0.161 / 0.410 / 4.6} \\
\midrule\midrule
Runtime top-$n$ HE-PLDA \added{(necessary)} & \numprint{16}\,s & \numprint{32}\,s & \numprint{48}\,s & \numprint{64}\,s & \numprint{80}\,s & \numprint{96}\,s & \numprint{128}\,s \\
\midrule\midrule
HE-PLDA (z-norm) & \multicolumn{7}{c}{\numprint{3724.80}\,s (for all \numprint{11640} reference-cohort comparisons)}\\\midrule\midrule
GMW pruning (BK: \numprint{28.2975}\,s) & \numprint{156.583}\,s&  \numprint{177.229}\,s& \numprint{197.791}\,s& \numprint{220.220}\,s& \numprint{247.070}\,s& \numprint{268.772}\,s& \numprint{282.889}\,s \\\midrule
improvement (az-norm) & \bf \numprint{18.5423672282775}$\times$ & \numprint{15.681618682547}$\times$ & \numprint{13.5897711870436}$\times$ & \numprint{11.918692553217}$\times$ & \numprint{10.4815437540011}$\times$ & \numprint{9.47618678121808}$\times$ & \numprint{8.48113500756512}$\times$ \\
\midrule\midrule
HE-PLDA (t-norm) & \multicolumn{7}{c}{\numprint{1219.84}\,s (for all \numprint{3812} cohort-probe comparisons)}\\\midrule\midrule
GMW pruning (BK: \numprint{16.8592}\,s) & \numprint{51.572}\,s & \numprint{58.718}\,s & \numprint{65.667}\,s & \numprint{72.550}\,s & \numprint{82.047}\,s & \numprint{89.239}\,s & \numprint{93.555}\,s \\\midrule
improvement (at-norm) & \bf \numprint{14.4477396981211}$\times$ & \numprint{11.3392057052981}$\times$ & \numprint{9.34555667750995}$\times$ & \numprint{7.95154397519836}$\times$ & \numprint{6.81832155621214}$\times$ & \numprint{6.03587760801432}$\times$ & \numprint{5.11647376708267}$\times$ \\ 
\midrule\midrule
\begin{tabular}{@{}c@{\hspace{2em}}c@{}}  
    \multirow[t]{2}{*}{
    \begin{tabular}{@{}c@{}}
         conventional \\ \added{(unprotected)} as-norm \\ 
    \end{tabular}} & \multirow[t]{2}{*}{
    \begin{tabular}{@{}r@{}}
         $C_{\text{llr}}^{\text{min}}$ \\
         minDCF \\
         EER
    \end{tabular}
}
\end{tabular} & \begin{tabular}{@{}c@{}} 0.161 \\ 0.390 \\ 4.6 \end{tabular} 
        & \begin{tabular}{@{}c@{}} 0.157 \\ 0.376 \\ 4.5 \end{tabular} 
        & \begin{tabular}{@{}c@{}} 0.156 \\ 0.374 \\ 4.5 \end{tabular} 
        & \begin{tabular}{@{}c@{}} 0.155 \\ 0.373 \\ 4.5 \end{tabular} 
        & \begin{tabular}{@{}c@{}} 0.155 \\ 0.374 \\ 4.4 \end{tabular} 
        & \begin{tabular}{@{}c@{}} 0.155 \\ \bf 0.372 \\ 4.4 \end{tabular} 
        & \begin{tabular}{@{}c@{}} 0.155 \\ 0.369 \\ 4.4 \end{tabular} \\
\midrule
\begin{tabular}{@{}c@{\hspace{2.75em}}c@{}} 
    \multirow[t]{2}{*}{
    \begin{tabular}{@{}c@{}}
     proposed \\
    \end{tabular}} & \multirow[t]{2}{*}{
    \begin{tabular}{@{}r@{}}
         $C_{\text{llr}}^{\text{min}}$ \\
         minDCF \\
         EER
    \end{tabular}
}
\end{tabular} & \begin{tabular}{@{}c@{}} 0.158 \\ 0.509 \\ 4.4 \end{tabular} 
         & \begin{tabular}{@{}c@{}} 0.151 \\ 0.492 \\ 4.3 \end{tabular} 
         & \begin{tabular}{@{}c@{}} 0.149 \\ 0.466 \\ 4.3 \end{tabular} 
         & \begin{tabular}{@{}c@{}} \bf 0.147 \\ 0.452 \\ \bf 4.1 \end{tabular} 
         & \begin{tabular}{@{}c@{}} 0.149  \\ 0.435 \\ 4.2 \end{tabular} 
         & \begin{tabular}{@{}c@{}} 0.149 \\ 0.429 \\ \bf 4.1 \end{tabular} 
         & \begin{tabular}{@{}c@{}} 0.149 \\ 0.408 \\ 4.2 \end{tabular} \\
\bottomrule
\end{tabular}}
\vspace{-0.5cm}
\end{table}

\section{Conclusions}
\label{sec:conclusion}

This paper reports the first approach to computationally manageable \added{(yet demanding)} privacy-preserving speaker recognition with cohort score normalisation.  Prior to this work, the latter was a computational bottleneck for PLDA with Paillier homomorphic encryption, with normalisation strategies that require many thousands of biometric comparisons being computationally prohibitive when performed in the encrypted domain. The set of cohort data used for score normalisation is pruned using a native binary speaker representation.  
%
Privacy is outsourced via secure multi-party computation through which a top-$n$ cohort set is pruned securely.  Privacy-insensitive cohort scores can then be decrypted and treated in the usual way.  
The cohort list is revealed to the sites capturing the reference and the probe data, respectively. This could be used by a security (\emph{not privacy}) adversary to mount hill-climbing attacks; instead, if the top-$n$ lists are in the province of the biometric service owner, these top-$n$ indices serve the intended recognition purpose. 
Future work could investigate the use of \deleted{Yao's garbled circuits and arithmetic} SMPC protocols for carrying out ranking and matrix operations in the protected domain (including PLDA) demanding less computational but more server communication overheads.


\vspace{-0.75em}
\paragraph*{Acknowledgements.} 
This work was supported by the BMBF and the HMWK within CRISP, the DFG as part of project E4 within the CRC 1119 CROSSING and A.1 within RTG 2050, by Omilia -- Conversational Intelligence, and by the Voice Personae and RESPECT projects, both funded by the French ANR.

\bibliographystyle{IEEEtran}
\bibliography{main}
\end{document}